\documentclass[12pt, notitlepage,aps,preprintnumbers,superscriptaddress,nofootinbib,aps,prd,floatfix]{revtex4}
\usepackage{array}
\usepackage{graphicx} 
\usepackage{dcolumn} 
\usepackage{orcidlink}
\usepackage{float}
\usepackage{physics,slashed}
\usepackage{amsfonts,amsmath,amssymb}
\usepackage{mathrsfs}
\usepackage{bm}

\usepackage{epsfig}
\usepackage{graphicx}   
\usepackage{subfigure}  
\usepackage{verbatim}   
\usepackage{color}      
\usepackage{hyperref}   
\usepackage{url}
\definecolor{nicered}{rgb}{0.5,0.,0.}
\definecolor{nicegreen}{rgb}{0.,0.5,0.}
\definecolor{niceblue}{rgb}{0.,0.,0.5}
\hypersetup{colorlinks,citecolor=nicegreen,linkcolor=nicered,urlcolor=niceblue}
\usepackage{array}
\usepackage{dcolumn}
\usepackage{multirow}
\usepackage{cprotect}
\usepackage{framed}
\usepackage{setspace}
\usepackage{enumitem}
\setlist{nolistsep}
\usepackage{mathtools}
\usepackage{tikz-feynman,contour}
\tikzfeynmanset{compat=1.1.0}
\usepackage{tikzsymbols}

\renewcommand\[{\begin{equation}}
\renewcommand\]{\end{equation}}
\begin{document}
\title{Explaining Fermions Mass and Mixing Hierarchies through $U(1)_X$ and $Z_2$ Symmetries}
\author{Abdul Rahaman Shaikh \orcidlink{0000-0003-2295-027X}\thanks{corresponding author}}
\email{abdulrahaman@ctp-jamia.res.in}
\author{Rathin Adhikari \orcidlink{0000-0002-8764-9587}}
\email{rathin@ctp-jamia.res.in}
\affiliation{Centre for Theoretical Physics, Jamia Millia Islamia (Central University),
Jamia Nagar, New Delhi-110025, India}
\date{\today} 
\begin{abstract}
For understanding the hierarchies of fermion masses and mixing, we extend the Standard Model gauge group with \( U(1)_X \) and \( Z_2 \) symmetry. The field content of the Standard Model is augmented by three heavy right-handed neutrinos, two new scalar singlets, and a scalar doublet. \( U(1)_X \) charges of different fields are determined after satisfying anomaly cancellation conditions. In this scenario, the fermion masses are generated through higher-dimensional effective operators with \( O(1) \) Yukawa couplings. The small neutrino masses are obtained through type-1 seesaw mechanism using the heavy right-handed neutrino fields, whose masses are generated by the new scalar fields. We discuss the flavour-changing neutral current processes that arise due to the sequential nature of \( U(1)_X \) symmetry. We have written effective higher-dimensional operators in terms of renormalizable dimension-four operators by introducing vector-like fermions.

\hspace*{\fill} 


\end{abstract}

\maketitle


\section{Introduction}\label{introduction}
The Standard Model (SM) of particle physics is the most successful theory that explains how the fundamental particles of our universe, such as quarks and leptons, interact with each other through gauge bosons, and how these particles acquire masses through the Higgs mechanism.

In the SM, all the charged fermions are directly coupled to a single scalar doublet known as the Higgs. After spontaneous symmetry breaking, the Higgs acquires a vacuum expectation value ($vev$), and the charged fermions gain masses. Experimentally, we know that the masses of these particles vary over a wide range of magnitudes. The mass of each charged fermion is equal to the $vev$ of the Higgs multiplied by the corresponding Yukawa coupling of the particular fermion field with the Higgs. The different fermion masses are obtained by varying the Yukawa couplings. For example, the Yukawa coupling for the top quark differs from that of the electron by about $10^6$. Neutrino masses are expected to be even smaller, of the order of $0.1$ eV. If one aims to generate neutrino masses in the SM scenario, similar to charged fermions, light right-handed neutrinos field is required to be introduced. However, the corresponding Yukawa couplings for neutrinos with the Higgs would need to be much smaller, about $10^{-12}$ times the Yukawa coupling of the top quark. There is no explanation for such a wide range of variation in Yukawa couplings within the SM framework.

Additionally, the mixing between the 1st and 2nd generations of quarks is large, while the mixing between the 2nd and 3rd generations, as well as between the 1st and 3rd generations, is very small. There is no explanation for this mixing pattern within the SM. A complete flavour theory should address all these issues, collectively known as the ``flavour puzzle" of the SM. For a basic introduction to flavour physics and mass matrix models, see, for example, \cite{Fritzsch:1977vd,Grossman:2010gw,Nomura:2023kwz, Gedalia:2010rj, Leurer:1992wg, Giudice:2008uua, Babu:2023zni, Abbas:2023ivi, Abbas:2022zfb}.

Flavour problems for charged fermions with different symmetries involving various fields have been studied by different authors \cite{PhysRevD.7.2457, Babu:1999me, PhysRevD.79.075014, Botella:2016krk, Antonelli:2009ws, Adhikari:2015woo}. Later, in \cite{Ghosh:2011up,Ganguly:2022cbo, Liao:2013saa, Ross:2002fb}, the authors explain neutrino masses and mixing. This type of work, involving different types of Abelian flavour symmetries, has been studied in various contexts \cite{VanLoi:2023utt, Ma_2021, Ma:2016zod, Rathsman:2023cic, Grossmann:2010ea, Asadi:2023ucx, Huitu:2017ukq, Fedele:2020fvh, Ma:2013mga, Kownacki_2017, FernandezNavarro:2023rhv, Mohanta:2022seo}. Apart from Abelian symmetries, there are other approaches to explain the hierarchies of fermion masses. These include the left-right symmetric model \cite{Ma:2020wjc, Bonilla:2023wok}, non-Abelian gauge symmetries \cite{2024EPJC...84..213G, Mohanta:2023soi}, hierarchical fermion wave functions \cite{Davidson:2007si}, strong dark-technicolor dynamics, multi-fermion condensates \cite{Abbas:2017vws, Abbas:2020frs}, discrete symmetries \cite{Abbas:2018lga, Abbas:2023ion, King:2013hoa}, and so on.

The smallness of the Yukawa couplings of fermions, except for the top quark, suggests that they may not directly interact with the Higgs, and their masses may not be directly linked to the Higgs $vev$. These small Yukawa couplings could be explained by introducing higher-dimensional operators involving new scalar fields. These operators would have an inverse power of mass dimension, making the corresponding dimensionless Yukawa couplings of $\mathcal{O}(1)$. This mechanism is known as the Froggatt-Nielsen (FN) mechanism \cite{Froggatt:1978nt}. 

The main idea of the FN mechanism is that different fermions carry different charges under a new flavour symmetry. The difference in FN charge in the Yukawa interaction is saturated by a number of insertions of new scalar fields known as flavon fields. Once the new flavon fields acquire $vev$, this new symmetry is broken, and the effective higher-dimensional operators restore the SM interactions, multiplied by some power of a small parameter $\epsilon$, which is equal to the $vev$ of the new scalar divided by the cutoff scale of the new theory. The hierarchy in the SM Yukawa couplings is then given by the different powers of $\epsilon$, and all the Yukawa couplings become $\mathcal{O}(1)$.

In this work, we extended the fermion sector of the SM by adding three heavy right-handed neutrinos, which are necessary for generating small neutrino masses. We also added an extra $U(1)_X$ symmetry to the SM symmetry group. Using the anomaly cancellation conditions \cite{Peskin:1995ev,Ma:2016zod}, we assigned charges to the fermions in such a way that only the top quark has a direct coupling with the SM Higgs, while direct couplings between all the light fermions and the Higgs are forbidden by the new flavour symmetry. This anomaly-free extra $U(1)_X$ symmetry naturally explains the inter-generational mass hierarchy of quarks and charged leptons. However, considering only the $U(1)_X$ symmetry explains only the inter-generational hierarchies of fermions. To account for the intra-generational mass hierarchies of fermion doublets, we need to introduce a $Z_2$ symmetry along with a new scalar singlet. The discrete $Z_2$ symmetry is widely used in different contexts of model building beyond the SM, such as in the two-Higgs-doublet model \cite{Branco:2011iw} and the minimal super-symmetric SM \cite{Csaki:1996ks}.

In our model, the Majorana neutrino masses are related to the $U(1)_X$ symmetry breaking scale and are proportional to the $vevs$ of the new singlet scalars. There is scope for heavy right-handed neutrinos having mass around $\sim$ TeV scale. Signatures of such right-handed neutrinos could be observable in near-future experiments. To obtain the observed small neutrino masses using type-I seesaw mechanism, the Dirac masses of the neutrinos need to be around $\sim$ MeV. We achieve this by introducing a new scalar doublet with a small $vev$ that couples only to neutrinos due to the conservation of a global lepton number symmetry \cite{Ma:2000cc}. The soft breaking of this global symmetry automatically results in the small $vev$ of this new scalar doublet. In this model, tree-level flavour-changing neutral current (FCNC) effects are predicted, but they are within experimental bounds.

In section \ref{model}, we discuss the details of the model with $U(1)_X$ and $Z_2$ symmetry, along with various field contents and their charges based on the anomaly cancellation conditions. The allowed higher-dimensional operators, based on the symmetries, are also discussed. In section \ref{fermion}, we calculate the masses and mixings of quarks and charged leptons and hence, discuss the CKM matrix for quarks. In section \ref{neutrino}, we calculate the small neutrino masses and their mixing using type-I seesaw mechanism. In section \ref{potential}, we discuss the scalar potential of our model. In section \ref{pheno}, we discuss the phenomenological implications of the various decay channels of the new gauge boson $Z'$, as well as various rare decays through $Z'$. In section \ref{dim4}, we discuss a possible ultra-violet (UV) complete theory by introducing some vector-like fermions with their charge assignments. In section \ref{benchmark}, we calculate benchmark values of $O(1)$ Yukawa couplings for the fermions to fit their masses and mixing. In section \ref{conclusion}, we present concluding remarks.
\section{The Model and Formalism}\label{model}
If we normalize the top quark mass as 1, then the masses of all other charged fermions, as well as the mixing angles of quarks, can be expressed in terms of a small parameter $\epsilon$. For example, if we consider $0.02 < \epsilon < 0.03$, the masses of quarks and charged leptons, along with the quark mixing angles, can be written as powers of $\epsilon$:
\begin{align}\label{supression of parameters}
     m_t &\approx 1 ~, & m_b &\approx \epsilon ~, & m_s &\approx \epsilon ~, & m_s &\approx \epsilon^2 ~, & m_u &\approx \epsilon^3 ~, & m_d &\approx \epsilon^3 ~, \nonumber \\ 
     m_{\tau} &\approx  \epsilon ~, & m_{\mu} &\approx \epsilon^2 ~, & m_e &\approx \epsilon^3 ~, &  s_{12}^q &\approx  \epsilon ~, & s_{23}^q &\approx \epsilon ~, & s_{13}^q &\approx \epsilon^2 ~,
\end{align} 
where $s_{ij}^{q} = \sin \theta_{ij}^{q} $ and $\theta_{ij}^{q}$ is the mixing angle between the $i^{\text{th}}$ and $j^{\text{th}}$ flavors of quarks. We can generate this kind of suppression in the masses of quarks and charged leptons, as well as the mixing angles of quarks, if the mass matrices of up-type, down-type quarks, and charged leptons take the following form :
\begin{align}\label{matrixstucture}
     m_u &\approx
    \begin{pmatrix}
        \epsilon^3 & \epsilon & 1 \\
        \epsilon^3 & \epsilon & 1 \\
        \epsilon^3 & \epsilon & 1 \\
    \end{pmatrix} ~,
&
    m_d &\approx
    \begin{pmatrix}
         \epsilon^3 & \epsilon^2  & \epsilon \\
         \epsilon^3 & \epsilon^2 & \epsilon \\
         \epsilon^3 & \epsilon^2 & \epsilon \\
    \end{pmatrix} ~,
&
    m_l &\approx
    \begin{pmatrix}
         \epsilon^3 & \epsilon^2  & \epsilon \\
         \epsilon^3 & \epsilon^2 & \epsilon \\
         \epsilon^3 & \epsilon^2 & \epsilon \\
    \end{pmatrix} ~. 
\end{align}
To obtain mass from these mass matrices, we have to diagonalize them. Since these matrices are not hermitian, we must diagonalize them using bi-unitary transformations. Following the notation of \cite{Rasin:1998je}, we can write 
\begin{align}\label{biunitary}
    m_D &= S^{\dag} m R ~,
    &
    m_D^2 &= S^{\dag} m m^{\dag} S ~,
    & 
    m_D^2 &= R^{\dag} m^{\dag} m R ~,
\end{align}
where $m_D$ is the diagonalized matrix corresponding to $m$, and $S$ and $R$ are the two unitary matrices. In other words, we can say that the eigenvalues of $m^\dagger m$ or $m m^\dagger$ give the mass squared of the fermions. The physical mixing matrices for quarks and leptons are the CKM and PMNS matrices, given by :
\begin{align}\label{ckm-pmns}
 V_{CKM} &= R^{u \dag} R^{d} ~,
 & V_{PMNS} &= R^{\nu \dag} R^{l} ~,
\end{align}
where $R^u$, $R^d$, $R^l$, and $R^{\nu}$ correspond to the $R$ matrix in Eq.~\eqref{biunitary} for $m$ corresponding to up-type quarks, down-type quarks, charged leptons and neutrinos, respectively. The CKM and PMNS matrices can be parameterized by three angles and one phase in the standard parameterization as in \cite{ParticleDataGroup:2020ssz} :
\begin{equation}\label{standard-ckm}
    V_{CKM} /V_{PMNS}=
    \begin{pmatrix}
       c_{12} c_{13} & s_{12} c_{13} & s_{13}~\textrm{e}^{\textrm{-i}\delta}\\
       -s_{12} c_{23} - c_{12} s_{23} s_{13}~\textrm{e}^{\textrm{i} \delta} & c_{12} c_{23} - s_{12} s_{23} s_{13} ~\textrm{e}^{\textrm{i} \delta} & s_{23} c_{13}\\
       s_{12} s_{23} - c_{12} c_{23} s_{13} ~\textrm{e}^{\textrm{i} \delta} & -c_{12} s_{23} - s_{12} c_{23} s_{13} ~\textrm{e}^{\textrm{i} \delta} & c_{23} c_{13}
    \end{pmatrix} ~,
\end{equation}
where $s_{ij} = \sin\theta_{ij}$ and $c_{ij} = \cos\theta_{ij}$, and $\delta$ is the CP-violating phase and Majorana phases of PMNS matrix are not shown here. Our main goal is to explain the hierarchy of fermion masses and mixing. The primary motivation for this work is to generate a kind of mass matrix, as shown in Eq.~\eqref{matrixstucture}, through a symmetry principle in a minimal setup. Let the $U(1)_X$ charges of the fermions be given by :
\begin{align}\label{chargedef}
    Q_{iL} &\rightarrow n_{1}^{i}~, & u_{iR} &\rightarrow n_{2}^{i}~, & d_{iR} &\rightarrow n_{3}^{i}~, & L_{iL} &\rightarrow n_{4}^{i}~, & e_{iR} &\rightarrow n_{5}^{i}~, & N_{iR} &\rightarrow n_{6} ^{i} ~~~,
\end{align}
where $Q_{iL} = (u,d)_{iL}$ are left-handed quark doublets ($i=1,2,3$ for the 1st, 2nd, and 3rd generations, respectively), $u_{iR}$ are the right-handed up-type quarks, $d_{iR}$ are the right-handed down-type quarks, $L_{iL} = (\nu,e)_{iL}$ are left-handed lepton doublets, $e_{iR}$ are the right-handed charged leptons and $N_{iR}$ are the right-handed heavy neutrinos. The anomaly cancellation conditions are :
\begin{align}
     [U(1)_Y]^2 U(1)_X : \sum_{i} 1/6 ~ n_{1}^{i} - 4/3~ n_{2}^i - 1/3~ n_3^i + 1/2~ n_4^i -n_5^i  &= 0 ~,
 \nonumber\\
     [U(1)_X]^2 U(1)_Y : \sum_{i}{n_1^i}^2 - 2 ~{n_2^i}^2 + {n_3^i}^2 - {n_4^i}^2 + {n_5^i}^2 &= 0 ~,
 \nonumber\\ 
     [SU(3)_c]^2 U(1)_X : \sum_{i} 2~ n_1^i - n_2^i - n_3^i &= 0 ~, \nonumber \\ 
    [SU(2)_L]^2 U(1)_X : \sum_{i}  9/2~ n_1^i + 3/2~ n_4 ^i &= 0 ~, \nonumber \\ 
     U(1)_X : \sum_{i} 6~ n_1^i - 3 ~ n_2^i - 3 ~ n_3^i + 2 ~ n_4^i - n_5^i- n_6^i &= 0 ~,  \nonumber\\ 
     [U(1)_X]^3 : \sum_{i} 6 ~{n_1^i}^3 - 3~ {n_2^i}^3 - 3 ~{n_3^i}^3 + 2~ {n_4^i}^3 - {n_5^i}^3 - {n_6^i}^3 &= 0 ~~.
\end{align}
It is noteworthy that, in the SM, the charges associated with different generations of fermions are identical. However, in our model, these charges are taken to differ across generations to establish mass hierarchies. This choice allows for anomaly cancellation to be managed generation by generation, which avoids the need for more complex solutions that might hinder achieving the desired mass hierarchies. Among the six anomaly cancellation conditions, only four are independent, so we can freely select two of the six charges. Using these conditions, the various \( U(1)_X \) charges can be expressed in terms of \( n_1 \) and \( n_2 \) as follows :
\begin{align}\label{anomalycondition}
  n_3^i &= 2 n_1^i - n_2^i ~,
&
  n_4^i &= - 3 n_1^i  ~,
 &  n_5^i &= -(2 n_1^i + n_2^i) ~,
 & n_6^i &= (n_2^i - 4 n_1^i) ~.
\end{align}
To create the matrix structure as shown in Eq.~\eqref{matrixstucture}, we need to assign sequential $U(1)_X$ charges. If we take $(n_1^1,n_1^2,n_1^3) = (0,0,0)$ and $(n_2^1,n_2^2,n_2^3) = (3,1,0)$ for the 1st, 2nd, and 3rd generations, respectively, then the charges of the other fields are :
\begin{align}\label{u1charges}
    (n_3^1,n_3^2,n_3^3) &= (-3,-1,0) ~,
 & (n_4^1,n_4^2,n_4^3) &= (0,0,0) ~, \nonumber\\
     (n_5^1,n_5^2,n_5^3) &= (-3,-1,0) ~,
 & (n_6^1,n_6^2,n_6^3) &= (3,1,0) ~.
\end{align}
If the scalar sector our model, includes one Higgs doublet and a scalar singlet \( \chi_1 \), with \( U(1)_X \) charges of 0 and -1, respectively, then only the third generation of quarks and charged lepton obtain masses from renormalizable dimension-4 operators through the Higgs doublet \( \phi \). Also, in this case, the second and first generations of quarks and charged leptons receive masses from dimension-5 and dimension-7 operators, respectively, via the scalar singlet \( \chi_1 \). Although this structure naturally generates a hierarchical mass pattern among the generations of charged fermions. However, hierarchy of masses in up-type and down-type quarks and charged leptons can not be obtained. To address the hierarchy, we introduce an additional scalar singlet \( \chi_2 \), which couples only to the second and third generations of down-type quarks and charged leptons. This coupling structure is enforced by adding a \( Z_2 \) symmetry with suitable charge assignments. As a result, the first generation of all charged fermion masses are obtained from a dimension-7 operator, while the second and third generations of up-type quark masses are obtained from dimension-5 and dimension-4 operators, respectively. Likewise, the second and third generations of down-type quark and charged lepton masses are obtained through dimension-6 and dimension-5 operators, respectively, as described later in Eq.~\eqref{eftlag} in the next section. The neutrino masses and mixing can be explained by type-I seesaw mechanism, as discussed in section \ref{neutrino}. Table~\ref{tab:1} summarizes the charge assignments for all particles under the various symmetries in our model.

\begin{table}[ht]
 \caption {Charges of scalar and fermion fields}
   \centering
    \begin{tabular}{m{3cm} m{1.5cm} m{1.5cm} m{1.5cm} m{2.5cm} m{2.5cm} m{1.5cm}}
        \hline
        Particles & $SU(3)_C$ & $SU(2)_L$ & $U(1)_Y$ & $U(1)_X$ & $Z_2$ & $U(1)_{L}$ \\
        \hline
        $Q_{iL}=(u,d)_{iL}$ &  3 & 2 & 1/6 &  (0, 0, 0) & (+, +, +) & 0\\
        $u_{iR}$ & 3 & 1 & 2/3 &  (3, 1, 0) & (+, +, +) & 0 \\
        $d_{iR}$ & 3 & 1 & - 1/3 &  (-3, -1, 0) & (+, -, -) & 0  \\
        $L_{iL}=(\nu ,l)_{iL}$ & 1 & 2 & -1/2 & (0,0,0) & (+, +, +)  & 1 \\
       $e_{iR}$ & 1 & 1 & -1 & (-3, -1, 0) & (+, -, -) & 1 \\
       $N_{iR}$ & 1 & 1 & 0 & (3, 1, 0) & (+, +, +) & 0 \\
       $\phi$ & 1 & 2 & 1/2 & 0 & + & 0 \\
       $\chi_1$ & 1 & 1 & 0 & -1 & +  & 0 \\
       $\chi_2$ & 1 & 1 & 0 & 0 & - & 0\\
       $\eta$ & 1 & 2 & 1/2 & 0 & + & -1 \\
       \hline
  \end{tabular}
    \label{tab:1}
\end{table}
We introduce a neutrino-philic scalar doublet, denoted by \( \eta \), to enable the neutrinos to acquire a Dirac mass independently of the SM Higgs doublet. The $vev$ of \( \eta \) is kept small by explicitly breaking the global lepton number symmetry $U(1)_L$, as elaborated in Section~\ref{potential}.
\section{Mass and Mixing of Charged Fermions} \label{fermion}
With respect to the charge assignment shown in Table~\ref{tab:1}, the Yukawa Lagrangian for charged fermions is given by :
\begin{multline} \label{eftlag}
     \mathcal{L}^{Y}_f = \left(\frac{\chi_1}{\Lambda}\right)^3 h^{u}_{i1} \bar{Q}_{iL} \tilde{\phi}\, u_{1R} +  \left(\frac{\chi_1}{\Lambda}\right) h^{u}_{i2} \bar{Q}_{iL} \tilde{\phi}\, u_{2R} + h^{u}_{i3} \bar{Q}_{iL} \tilde{\phi} \, u_{3R} +\left(\frac{\chi_1^{*}}{\Lambda}\right)^3 h^{d}_{i1} \bar{Q}_{iL} \phi \, d_{1R} \\
     +  \left(\frac{\chi_1^{*} \chi_2}{\Lambda^{2}}\right) h^{d}_{i2} 
     \bar{Q}_{iL}  \phi~ d_{2R}  + \left(\frac{\chi_2}{\Lambda} \right) h^{d}_{i3} \bar{Q}_{iL} \phi \, d_{3R} + \left(\frac{\chi_1^{*}}{\Lambda}\right)^3 h^{l}_{i1} \bar{L}_{1L} \phi \, e_{1R} + \left(\frac{\chi_1^{*} \chi_2}{\Lambda^{2} }\right) h^{l}_{i2} \\
     \bar{L}_{2L} \phi \, e_{2R}
     + \left(\frac{\chi_2}{\Lambda}\right) 
    h^{l}_{i3} \bar{L}_{3L} \phi \, e_{3R} +  \textrm{h.c} ~,
\end{multline}
where \( h_{ij} \) are the Yukawa couplings, and the superscripts \( u \), \( d \) and \( l \) refer to up-type quarks, down-type quarks and charged leptons, respectively. After the flavour symmetry breaking, the scalar fields \( \chi_1 \) and \( \chi_2 \) acquire $vevs$. If we define \( \frac{v_1}{\Lambda} = \epsilon \) and \( \frac{v_2}{\Lambda} = \epsilon' \), where \( v_1 \) and \( v_2 \) are the $vevs$ of \( \chi_1 \) and \( \chi_2 \), respectively, then the Yukawa Lagrangian for quarks is given by :
\begin{multline}\label{quarklag}
    \mathcal{L}^{Y}_{Q} = \epsilon^3 h^{u}_{i1} \bar{Q}_{iL} \tilde{\phi}~ u_{1R} + \epsilon h^{u}_{i2} \bar{Q}_{iL}\tilde{\phi} ~ u_{2R} +h^{u}_{i3} \bar{Q}_{iL} \tilde{\phi} ~ u_{3R}
    +\epsilon^3 h^{d}_{i1} \bar{Q}_{iL} \phi ~ d_{1R}
    + \epsilon \epsilon' h^{d}_{i2} \bar{Q}_{iL} \phi 
    ~ d_{2R}  \\
    + \epsilon'h^{d}_{i3} \bar{Q}_{iL} \phi ~ d_{3R} +  \textrm{h.c} ~.
\end{multline}
Then, the mass matrices of the up-type and down-type quarks become :
\begin{align}\label{matrixchargedfermions}
     m_u &=
    \begin{pmatrix}
        h^{u}_{11} \epsilon^3 & h^u_{12} \epsilon & h^u_{13} \\
         h^u_{21} \epsilon^3 & h^u_{22} \epsilon & h^u_{23} \\
          h^u_{31} \epsilon^3 & h^u_{32} \epsilon & h^u_{33} \\
    \end{pmatrix}
    \frac{v}{\sqrt{2}} ~,
&
    m_d &=
    \begin{pmatrix}
        h^{d}_{11} \epsilon^3 & h^d_{12} \epsilon \epsilon'  & h^d_{13} \epsilon' \\
         h^d_{21} \epsilon^3 & h^d_{22} \epsilon \epsilon' & h^d_{23}  \epsilon' \\
          h^d_{31} \epsilon^3 & h^d_{32} \epsilon \epsilon' & h^d_{33} \epsilon' \\
    \end{pmatrix}
     \frac{v}{\sqrt{2}} ~,
\end{align}
where \( v \) is the $vev$ of the Higgs field. We will obtain the masses of the quarks by diagonalizing the above mass matrices. These matrices are not hermitian, so we need to diagonlize following hermitian :
\begin{align}\label{quarksmasssquare}
    m^{\dag}_u m_u &=
    \begin{pmatrix}
        x^{u}_{11} \epsilon^6 & x^u_{12} \epsilon^4 & x^u_{13} \epsilon^3 \\
         x^u_{12} \epsilon^4 & x^u_{22} \epsilon^2 & x^u_{23} \epsilon \\
          x^u_{13} \epsilon^3 & x^u_{23} \epsilon & x^u_{33} \\
    \end{pmatrix} \frac{v^2}{2} ~,
&
    m^{\dag}_d m_d &=
    \begin{pmatrix}
        x^{d}_{11} \epsilon^6 & x^d_{12} \epsilon^4 \epsilon'  & x^d_{13} \epsilon^3 \epsilon'  \\
         x^d_{12} \epsilon^4 \epsilon'   & x^d_{22} \epsilon^2 \epsilon'^2 & x^d_{23} \epsilon \epsilon'^2 \\
          x^d_{13} \epsilon^3 \epsilon' & x^d_{23} \epsilon \epsilon'^2 & x^d_{33} \epsilon'^2 \\
    \end{pmatrix} \frac{v^2}{2} ~,
\end{align}
where
\begin{align}\label{notation}
    x_{11} &= h_{11}^2 + h_{21}^2 + h_{31}^2 ~,&
    x_{12} &=  h_{12} h_{11} + h_{22} h_{21} +  h_{32} h_{31} ~, \nonumber\\
   x_{22} &= h_{12}^2 + h_{22}^2 + h_{32}^2 ~, &
    x_{23} &=  h_{13} h_{12}  + h_{23} h_{22} +  h_{33} h_{32} ~, \nonumber\\
   x_{33} &= h_{13}^2 + h_{23}^2 + h_{33}^2 ~, &
   x_{13} &= h_{11} h_{13} + h_{21} h_{23} + h_{31} h_{33} ~.
    \end{align}
For up-type and down-type quarks, replace \( x \) with \( x^u \) and \( x^d \), respectively and replace \( h \) with \( h^u \) and \( h^d \), respectively. Following \cite{Rasin:1998je}, we obtained the masses of the quarks in the leading order of $\epsilon$'s as :
\begin{equation}\label{upmass}
  (m_u, m_c, m_t) \approx \left(\sqrt{\frac{x^u_{11} {x^u_{23}}^2 + {x^u_{12}}^2 x^u_{33} - x^u_{11} x^u_{22} x^u_{33}}{{x^u_{23}}^2 - x^u_{22} x^u_{33}}}  \epsilon^3,  \sqrt{x^u _{22} - \frac{{x^u_{23}}^2}{ x^u_{33}}} \epsilon,  \sqrt{x^u_{33}}\right)\frac{v}{\sqrt{2}} ~,
\end{equation}
\begin{equation}\label{downmass}
  (m_d, m_s, m_b) \approx \left(\sqrt{\frac{x^d_{11} {x^d_{23}}^2 + {x^d_{12}}^2 x^d_{33} - x^d_{11} x^d_{22} x^d_{33}}{{x^d_{23}}^2 - x^d_{22} x^d_{33}}}  \epsilon^3, \sqrt{x^d _{22} - \frac{{x^d_{23}}^2}{ x^d_{33}}}  \epsilon \epsilon',  \sqrt{x^d_{33}} \epsilon'\right)\frac{v}{\sqrt{2}} ~.
\end{equation}
The mixing angles of up-type quarks and down-type quarks in the leading order of $\epsilon$'s are given as :
\begin{equation}\label{upangle}
(s^u_{12},s^u_{23},s^u_{13}) \approx \left(\left| \frac{x^u_{13} x^u_{23}-x^u_{12} x^u_{33}}{{x^u_{23}}^2 - x^u_{22}x^u_{33}}\right| \epsilon^2, \frac{x^u_{23}}{x^u_{33}}\epsilon, \frac{x^u_{13}}{x^u_{33}} \epsilon^3 \right)~,
\end{equation}
\begin{equation}\label{downangle}
(s^d_{12},s^d_{23},s^d_{13}) \approx \left(\left|\frac{x^d_{13} x^d_{23}-x^d_{12} x^d_{33}}{{x^d_{23}}^2 - x^d_{22}x^d_{33}} \right|\frac{\epsilon^2}{\epsilon'}, \frac{x^d_{23}}{x^d_{33}}\epsilon, \frac{x^d_{13} }{x^d_{33} } \frac{\epsilon^3}{\epsilon'}\right) ~.
\end{equation}
We can write the mixing matrices for up type and down type quarks in the leading order  of $\epsilon$'s are :
\begin{align}\label{rotationquarks}
     R^u &\approx
    \begin{pmatrix}
       1 & s_{12}^u & s_{13}^u \\
       -s^u_{12} & 1 & s^u_{23} \\
      (s^u_{12} s^u_{23}- s^u_{13})& s^u_{23} & 1 \\
    \end{pmatrix} ~,
&
    R^d &\approx
    \begin{pmatrix}
         1 & s^d_{12}  & s^d_{13}  \\
       -s^d_{12}  & 1 & s^d_{23} \\
      (s^d_{12} s^d_{23} - s^d_{13})  & s^d_{23} & 1 \\
    \end{pmatrix} ~.
\end{align}
From Eq.~\eqref{ckm-pmns}, the CKM matrix in the leading order of $\epsilon$'s is given by :
\begin{equation}\label{ckmmatrix}
    V_{CKM} = R^{u \dag }.R^d \approx
    \begin{pmatrix}
         1 & s^d_{12}  & s^d_{13} \\
       -s^d_{12}  & 1 & (s^d_{23} + s^u_{23})\\
      (s^d_{12}( s^d_{23} + s^u_{23}) - s^d_{13})  & (s^d_{23} + s^u_{23}) & 1 \\
    \end{pmatrix} ~.
\end{equation}
By following the same approach, we can write the Yukawa interactions for charged leptons after the new scalar fields \( \chi_1 \) and \( \chi_2 \) acquire their $vevs$ as : 
\begin{equation}\label{chargedleptonlag}
    \mathcal{L}^{Y}_{l} = \epsilon^3 h^{l}_{i1} \bar{L}_{i L} \phi ~ e_{1R} + \epsilon \epsilon' h^{l}_{i2} \bar{L}_{i L} \phi ~ e_{2R} + \epsilon' h^{l}_{i3} \bar{L}_{i L} \phi ~ e_{3R}  + \textrm{h.c} ~~.
\end{equation}
Then we can write the mass matrix of charged leptons as :
\begin{align}\label{chargedleptonsmassmatrix}
   m_l &=
    \begin{pmatrix}
        h^{l}_{11} \epsilon^3 & h^l_{12} \epsilon \epsilon'  & h^l_{13} \epsilon'  \\
         h^l_{21} \epsilon^3 & h^l_{22} \epsilon \epsilon'  & h^l_{23} \epsilon'  \\
          h^l_{31} \epsilon^3 & h^l_{32} \epsilon \epsilon'  & h^l_{33} \epsilon' 
    \end{pmatrix}
     \frac{v}{\sqrt{2}}~,
&
    m^{\dag}_l m_l &=
    \begin{pmatrix}
        x^{l}_{11} \epsilon^6 & x^l_{12} \epsilon^4 \epsilon'  & x^l_{13} \epsilon^3 \epsilon'  \\
         x^l_{12} \epsilon^4 \epsilon'   & x^l_{22} \epsilon^2 \epsilon'^2 & x^l_{23} \epsilon \epsilon'^2 \\
          x^l_{13} \epsilon^3 \epsilon' & x^l_{23} \epsilon \epsilon'^2 & x^l_{33} \epsilon'^2 \\
    \end{pmatrix} \frac{v^2}{2} ~,
\end{align}
where $x^l_{ij}$ is same as defined in Eq.~\eqref{notation} with the replacement of $x_{ij}$ by $x^l_{ij}$ and $h_{ij}$ by $h_{ij}^l$.
After diagonalizing we get the masses and  mixing angles of charged leptons in the leading order of $\epsilon$'s as : 
\begin{equation}\label{chargedleptonmasses}
  (m_e, m_{\mu}, m_{\tau}) \approx \left(\sqrt{\frac{x^l_{11} {x^l_{23}}^2 + {x^l_{12}}^2 x^l_{33} - x^l_{11} x^l_{22} x^l_{33}}{{x^l_{23}}^2 - x^l_{22} x^l_{33}}}  \epsilon^3,  \sqrt{x^l _{22} - \frac{{x^l_{23}}^2}{ x^l_{33}}}  \epsilon \epsilon',  \sqrt{x^l_{33}} \epsilon'\right)\frac{v}{\sqrt{2}} ~,
\end{equation} 
\begin{equation}\label{leptonangle}
(s^l_{12},s^l_{23},s^l_{13}) \approx \left(\left|\frac{x^l_{13} x^l_{23}-x^l_{12} x^l_{33}}{{x^l_{23}}^2 - x^l_{22}x^l_{33}  } \right|\frac{\epsilon^2}{\epsilon^{'}}, \frac{x^l_{23}}{x^l_{33}}\epsilon, \frac{x^l_{13} }{x^l_{33} } \frac{\epsilon^3}{\epsilon^{'}}\right) ~.
\end{equation}
Then the rotation matrix for charged leptons in the leading order can be written as :
\begin{align}\label{rotationchargedleptons}
    R^l &=
    \begin{pmatrix}
         1 & s^l_{12}  & s^l_{13}  \\
       s^l_{12}  & 1 & s^l_{23} \\
      (s^l_{12} s^l_{23} - s^l_{13})  & s^l_{23} & 1 \\
    \end{pmatrix} ~.
\end{align}
\section{Mass and Mixing of Neutrinos}\label{neutrino}
Unlike the mass generation of up-types quarks, we need to consider another mechanism to explain the very small masses of the three active neutrinos. Here, we consider type-I seesaw mechanism to explain the small neutrino masses. We know that the mass of active neutrinos is at most on the order of 0.1 eV. In our model, the heavy right-handed neutrinos are charged under the new \( U(1)_X \) symmetry, so their masses are related to the symmetry-breaking scale of \( U(1)_X \). If this scale is required to be around 1 TeV, then the Dirac mass \( M_D \) in the seesaw mechanism should be on the order of \(\sim 1\) MeV to obtain the appropriate small light neutrino masses. We introduce a new scalar doublet that couples only to neutrinos. This can be achieved by assigning the lepton number of the heavy right-handed neutrino to \( L = 0 \), which prevents the Yukawa interaction of neutrinos with the SM Higgs. Additionally, we introduce a new scalar doublet \( \eta \) with a lepton number \( L = -1 \) \cite{Ma:2000cc}. The Yukawa Lagrangian for neutrinos can then be written as :
\begin{multline} \label{eftlagneutrions}
   \mathcal{L}^{Y}_{\nu} =  \left(\frac{\chi_1}{\Lambda}\right)^3 h^{\nu}_{i1} \bar{L}_{iL} \tilde{\eta} N_{1R} + \left(\frac{\chi_1}{\Lambda}\right) h^{\nu}_{i2} \bar{L}_{iL} \tilde{\eta} N_{2R} +  h^{\nu}_{i3} \bar{L}_{iL} \tilde{\eta} N_{3R}
    + \left(\frac{\chi_1}{\Lambda}\right)^5 h^m_{11}\, \chi_1\, \bar{N}_{1R}^c N_{1R}
    +\left(\frac{\chi_1}{\Lambda}\right)^3\\  h^m_{12}\, \chi_1 \,\bar{N}_{1R}^c N_{2R} 
   + \left(\frac{\chi_1}{\Lambda}\right)^2 h^m_{13} \,\chi_1 \,\bar{N}_{1R}^c N_{3R} + \left(\frac{\chi_1}{\Lambda}\right)^3 h^m_{21}\, \chi_1 \,\bar{N}_{2R}^c N_{1R} 
     + \left(\frac{\chi_1}{\Lambda}\right) h^m_{22} \,\chi_1 \,\bar{N}_{2R}^c N_{2R}
    \\ + h^m_{23}\, \chi_1\,  \bar{N}_{2R}^c N_{3R} + \left(\frac{\chi_1}{\Lambda}\right)^2 h^m_{31}\, \chi_1 \,\bar{N}_{3R}^c N_{1R} + h^m_{32}\, \chi_1\, \bar{N}_{2R}^c N_{3R} + h^m_{33} \,\chi_2^{*} \,\bar{N}_{3R}^c N_{3R} + \textrm{h.c} ~.
\end{multline}
The effective Yukawa interactions for neutrinos, after the new scalar singlets acquire $vevs$, take the form :
\begin{multline}\label{neutrinolag}
    \mathcal{L}^{Y}_{\nu} =  \epsilon^3 h^{\nu}_{i1} \bar{L}_{iL} \tilde{\eta} N_{1R} + \epsilon h^{\nu}_{i2} \bar{L}_{iL} \tilde{\eta} N_{2R} +  h^{\nu}_{i3} \bar{L}_{iL} \tilde{\eta} N_{3R} + \epsilon^5 v_1 \, h^m_{11} \bar{N}_{1R}^c N_{1R} + \epsilon^3 v_1 \,
     h^m_{12} \bar{N}_{1R}^c N_{2R}\\ 
     +\epsilon^2 v_1 \, h^m_{13}
     \bar{N}_{1R}^c N_{3R}
     + \epsilon^3 v_1 \, h^m_{21} \bar{N}_{2R}^c N_{1R}
     + \epsilon v_1 \,
     h^m_{22} \bar{N}_{2R}^c N_{2R} 
     + v_1 \, h^m_{23} \bar{N}_{2R}^c N_{3R}
     + \epsilon^2 v_1 \, h^m_{31} \bar{N}_{3R}^c N_{1R}\\
     +  v_1 \, h^m_{23} \bar{N}_{2R}^c N_{3R} + v_2 \,  h^m_{33} \bar{N}_{3R}^c N_{3R} + \textrm{h.c} ~.
\end{multline}
The corresponding Dirac and Majorana mass matrices for neutrinos are given by :
\begin{align} \label{neutrinomassmatrix}
   M_D &=
\begin{pmatrix}
    h^{\nu}_{11}\epsilon^3 & h^{\nu}_{12}\epsilon & h^{\nu}_{13}  \\
    h^{\nu}_{21}\epsilon^3 & h^{\nu}_{22} \epsilon & h^{\nu}_{23} \\
    h^{\nu}_{31} \epsilon^3 & h^{\nu}_{32} \epsilon & h^{\nu}_{33} 
\end{pmatrix} \frac{v_{\eta}}{\sqrt{2}} ~,
&
  M_R =
\begin{pmatrix}
    h^m_{11} \epsilon^5  & h^m_{12} \epsilon^3 & h^m_{13} \epsilon^2 \\
    h^m_{21} \epsilon^3 &  h^m_{22} \epsilon  & h^m_{23} \\
    h^m_{31} \epsilon^2 & h^m_{32} & h^m_{33} \, \epsilon'/\epsilon
\end{pmatrix}\frac{v_1}{\sqrt{2}} ~,
\end{align}
where $v_{\eta}$ is the $vev$  of $\eta$, and  $v_1$ and $v_2$ are $vevs$ of $\chi_1$ and $\chi_2$. Using the seesaw formula \cite{Yanagida:1979as,Minkowski:1977sc,Mohapatra:2004zh,Grimus:2000vj}, we obtain the light neutrino mass matrix as :
\begin{equation*}
M_{\nu} \approx - M_{D} M^{-1}_{R} {M_D}^T 
\end{equation*}
\begin{equation} \label{seesaw}
\approx -\frac{v_{\eta}^2 \epsilon}{\sqrt{2}v_{1}}
 \begin{pmatrix}
 \frac{(h_{11}^{\nu})^2 - 2 h_{11}^{\nu} + h_{11}^{m}}{h_{11}^{m} -1} & \frac{-1 +h_{11}^{\nu}  -h_{11}^{\nu} h_{22}^{\nu} +h_{22}^{\nu} h_{11}^{m}}{h_{11}^{m}-1} & 1 \\
\frac{-1 +h_{11}^{\nu}  -h_{11}^{\nu} h_{22}^{\nu} +h_{22}^{\nu} h_{11}^{m}}{h_{11}^{m}-1} & \frac{2-4 h_{22}^{\nu} + (h_{2}^{\nu})^2 -h_{11}^{m} + 2 h_{22}^{\nu} h_{11}^{m}}{h_{11}^{m}-1} & 1 - h^{\nu}_{33} + h_{22}^{\nu} h_{33}^{\nu} \\
1 & 1 - h^{\nu}_{33} + h_{22}^{\nu} h_{33}^{\nu} & h_{33}^{\nu} - (-1+h_{33}^{\nu})h_{33}^{\nu}
 \end{pmatrix} ~.
\end{equation} 
Here, we have taken all the off-diagonal Yukawa couplings of the Dirac mass matrix and the  Majorana mass matrix as well as  $h^{m}_{22}$, to be equal to 1. After diagonalizing this effective light neutrino mass matrix, we obtain the three light neutrino masses as :
\begin{equation*}
    m_{\nu 1} \approx \Big[ \frac{
    \substack{
        [h^{\nu}_{11} (h^{\nu}_{22} h^{\nu}_{33}-1)]^2 
        - 2 h^{\nu}_{11} (h^{\nu}_{22} h^{\nu}_{33} (2 h^{\nu}_{22}+ h^{\nu}_{33} -4) +1) 
        + h^{\nu}_{33}(2 (h^{\nu}_{22}-1) h^{\nu}_{22} h^m_{11} + h^{\nu}_{33} - 2) + h^m_{11}
    }}{
    h^{\nu}_{33} (h^{\nu}_{22} h^m_{11} (h^{\nu}_{22} h^{\nu}_{33}-2) 
    - 2 h^{\nu}_{22} (h^{\nu}_{22} + h^{\nu}_{33} - 3) + h^{\nu}_{33} - 2) + h^m_{11} - 1}\Big]
    \frac{v_{\eta}^2 \epsilon}{\sqrt{2}v_1} ~,
\end{equation*}
\begin{equation*} 
 m_{\nu 2} \approx  \Big[\frac{ (h^{\nu}_{33} (h^{\nu}_{22} h^m_{11} (h^{\nu}_{22} h^{\nu}_{33}-2)-2 h^{\nu}_{22} (h^{\nu}_{22}+h^{\nu}_{33}-3)+h^{\nu}_{33}-2)+h^m_{11}-1)}{h^{\nu}_{33}(h^{\nu}_{33}-2) (h^m_{11}-1)}\Big] \frac{v_{\eta}^2 \epsilon}{\sqrt{2}v_1} ~,
\end{equation*}
\begin{equation}\label{neutrinomass}
      m_{\nu 3} \approx \Big[(h^{\nu}_{33}-2) h^{\nu}_{33} \Big]\frac{v_{\eta}^2 \epsilon}{\sqrt{2}v_{1}} ~.
\end{equation}
The neutrino mixing angles are obtained  by following \cite{Abbas:2022zfb,Abbas:2023ion} as :
\begin{multline}\label{neutrinomixing}
(s_{12}^{\nu},s_{23}^{\nu},s_{13}^{\nu}) \approx \Bigl(\frac{-h^{\nu}_{11} h^{\nu}_{22}+ h^{\nu}_{11}+h^{\nu}_{22} h^m_{11}-1}{{h^{\nu}_{22}}^2+2 h^{\nu}_{22} h^m_{11}-4 h^{\nu}_{22}-h^m_{11}+2},\frac{h^{\nu}_{22} h^{\nu}_{33}-h^{\nu}_{33}+1}{h^{\nu}_{33}-(h^{\nu}_{33}-1) h^{\nu}_{33}}, \\
    \frac{h^{\nu}_{11} (h^{\nu}_{22}-1)-h^{\nu}_{22} h^m_{11}+1}{h^{\nu}_{22} (h^{\nu}_{22}+ 2 h^m_{11}-4)-h^m_{11}+2}+\frac{1}{ h^{\nu}_{33}(2 - {h^{\nu}_{33}})}
  \Bigr) ~.
\end{multline}
Since all off-diagonal elements of the charged lepton rotation matrix $R^l$ in Eq.~\eqref{rotationchargedleptons} are suppressed by at least one power of $\epsilon$'s, we can say that the rotation matrix corresponding to the charged leptons is almost an identity matrix. Therefore, the PMNS matrix becomes :
\begin{equation}\label{pmns}
    V_{PMNS} = R^{\nu \dag} R^{l} \approx R^{\nu \dag} ~.
\end{equation}
The masses in Eq.~\eqref{neutrinomass} and the mixing angles in Eq.~\eqref{neutrinomixing} can satisfy both the normal and inverted hierarchy of neutrino mass-squared differences and mixing angles with $\mathcal{O}(1)$ values of dimensionless Yukawa couplings.
\section{Scalar Potential}\label{potential}
As we discuss above for explaining the fermion masses with $O(1)$ Yukawa couplings, we need  the $vevs$ of the scalar doublets of our model to be hierarchical in nature and the $vevs$ of scalar singlets are higher than the SM Higgs $vev$. We can achieve this kind of $vevs$ of our model simply by adding some soft lepton number symmetry breaking term as in \cite{Ma:2000cc,Ma:2016zod}. Let's discuss the scalar potential of the models for generating those hierarchical $vevs$. The scalar potential of our model is given by
\begin{multline}
 V = \mu_{11}^2 \Phi^\dagger \Phi + \mu_{22}^2 \eta^\dagger \eta + 
\mu_{33}^2 \chi_1^\dagger \chi_1 + \mu_{44}^2 \chi_2^\dagger \chi_2  + \frac{1}{2} \lambda_{11} (\Phi^\dagger \Phi)^2 + \frac{1}{2} \lambda_{22} (\eta^\dagger \eta)^2 + \frac{1}{2} \lambda_{33} (\chi_1^\dagger \chi_1)^2 \\
+ \frac{1}{2} \lambda_{44} (\chi_2^\dagger \chi_2)^2  + \lambda_{12} (\Phi^\dagger \Phi)(\eta^\dagger \eta) + \lambda'_{12} (\Phi^\dagger \eta)(\eta^\dagger \phi)+ \lambda_{13} (\Phi^\dagger \Phi)(\chi_1^\dagger \chi_1)  + 
\lambda_{14} (\Phi^\dagger \Phi)(\chi_2^\dagger \chi_2)\\
+\lambda_{23} (\eta^\dagger \eta)(\chi_1^\dagger \chi_1) 
+ \lambda_{24} (\eta^\dagger \eta)(\chi_2^\dagger \chi_2) + \mu_{12}^2(\Phi^\dagger \eta + \eta^\dagger \Phi) ~,
\end{multline}
where $\mu_{12}^2$ is the soft symmetry breaking term which break lepton number symmetry. Let $vevs$ of $\phi,\eta,\chi_1 ,\chi_2$ be respectively $v,v_{\eta},v_1,v_2$. If we consider $\mu_{11}^2$, $\mu_{33}^2$ and $\mu_{44}^2$ as positive and $\mu_{22}^2$ as negative and $\mu_{12}^2 << \mu_{22}^2$, then with the minimization of the scalar potential, the constrained equations involving different couplings are as follows :  
\begin{eqnarray}\label{constrained}
&& \mu_{11}^2 + \lambda_{11} v^2 + \lambda_{13} v_1^2 +\lambda_{14} v_2^2  \approx  0 ~~, \\ 
&& \mu_{33}^2 + \lambda_{33} v_1^2 + \lambda_{13} v^2 \approx 0 ~~, \\
&& \mu_{44}^2 + \lambda_{44} v_2^2 + \lambda_{14} v^2  \approx  0  ~~, \\
&& v_{\eta} \approx \frac{- \mu_{12}^2 v}{\mu_{22}^2+(\lambda_{12}+\lambda'_{12})v^2+\lambda_{23}v_1^2+\lambda_{24}v_2^2} ~~.
\end{eqnarray}
As $\mu_{12}^2 << \mu_{22}^2$, $v_{\eta}$ automatically becomes small compared to $v$ and by adjusting the unknown parameters, $\lambda_{11},\lambda_{13},\lambda_{14},\lambda_{33},\lambda_{13}$, $\mu_{11}^2$ and $\mu_{33}^2$, it can be possible for the $vevs$ of $\chi_1$ and $\chi_2$ to be greater than $v$.

The singlets are purely responsible for the breaking of the new symmetry and giving mass to the heavy right-handed neutrinos. We have considered negligibly small couplings between the scalar doublets and the singlets. From the two Higgs doublets, there will be five physical scalars: two neutral scalars, two charged scalars, and one pseudo-scalar. One of the neutral scalars behaves like the SM Higgs boson, while the other scalars have masses approximately \( \sim \mu_{22} \), which should be at the TeV scale to achieve the required hierarchy between the $vevs$ of the Higgs doublets and these scalars may be accessible in future accelerators as discussed in \cite{Ma:2000cc}. To ensure vacuum stability at the tree level, the couplings should satisfy \( \lambda_{11,22} > 0 \), \( \lambda_{12} > - (\lambda_{11} \lambda_{22})^{1/2} \), and \( \lambda_{12} + \lambda'_{12} > - (\lambda_{11} \lambda_{22})^{1/2} \) \cite{Haba:2011fn,Machado:2015sha}. The quartic couplings cannot be too large, as this could violate unitarity in tree-level scalar-scalar scattering at sufficiently high scales, which is known as the perturbative unitarity condition. In our model this condition can be achieved by considering all \( |\lambda_{ij}| \) values smaller than \( 8 \pi \), as discussed in \cite{Machado:2015sha,Kanemura:2015ska,Goodsell:2018fex}.
\section{Phenomenology of the model} \label{pheno} 
In our model, all fermions couple to a single Higgs doublet, enabling simultaneous diagonalization of the Yukawa and fermion mass matrices, thereby preventing tree-level scalar-mediated FCNCs \cite{Sher:2022aaa,Glashow:1976nt}. While FCNCs may arise at the one-loop level, the presence of an additional scalar doublet also influences oblique parameters. The effects of a second Higgs doublet on Electroweak Precision Tests (EWPT), particularly on oblique parameters and various FCNC processes, are extensively studied in \cite{Machado:2015sha}, with detailed phenomenological analyses under collider constraints available in \cite{Camargo:2018klg}. Due to the neutrino-philic nature of the second Higgs doublet and its minimal overlap with Standard Model fields, the charged scalars predominantly decay into neutrinos. Consequently, LEP constraints require their masses to exceed \(80\)~GeV, as discussed in \cite{Machado:2015sha}.

\begin{figure}[ht]
    \centering
\includegraphics{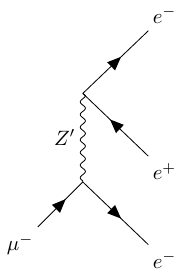}
    \caption{Feynman diagram for $\mu \rightarrow 3 e $ decay at tree-level through new gauge boson.}
    \label{fig:muondecay}
\end{figure}

\begin{figure}[ht]
    \centering
\includegraphics[scale=0.7]{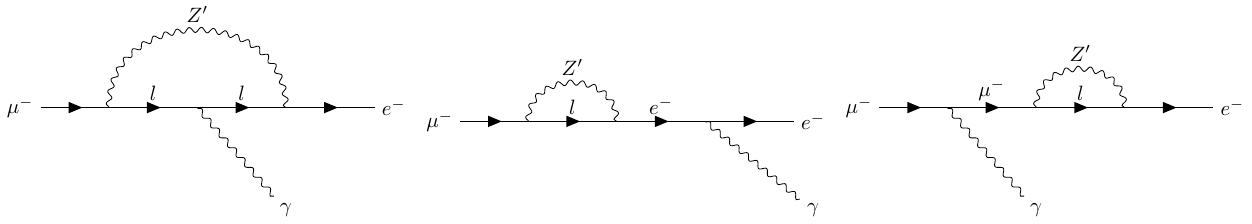}
    \caption{Feynman diagrams for $\mu \rightarrow e \gamma $ at one-loop level through new gauge boson,$Z'$.}
    \label{fig:mutoeggma}
\end{figure}

The different generations of up and down-type quarks and charged leptons have different \( U(1)_X \) charges, leading to the possibility of FCNCs mediated by the new gauge boson \( Z' \). There will be no mass mixing between \( Z \) and \( Z' \), as there is no common \( vev \) that spontaneously breaks both \( U(1)_X \) and \( SU(2)_L \times U(1)_Y \) \cite{Langacker:2008yv,Adhikari:2008uc}. Furthermore, we assume that the kinetic mixing between \( Z \) and \( Z' \) is negligible. Since \( Z' \) couples to quarks and leptons according to the \( U(1)_X \) charges given in Table~\ref{tab:1}, the branching ratios of \( Z' \rightarrow e^- e^+ \) and \( Z' \rightarrow \mu^- \mu^+ \) are proportional to $(n_5^{1,2})^2/\sum_{i=1}^3 3 ((n_2^i)^2 +(n_3^i)^2) + ((n_5^i)^2 +(n_6^i)^2$ and their ratio is given as :
\begin{equation}
    \frac{\Gamma (Z' \rightarrow e^{-} e^{+})}{\Gamma (Z' \rightarrow \mu^{-} \mu^{+})} = \frac{(n_5^1)^2}{(n_5^2)^2} = 9 ~~.
\end{equation}
This ratio can be used to distinguish this model from other \( U(1)_X \) models. One interesting feature is that the \( \tau \) lepton has zero \( U(1)_X \) charge, so the \( Z' \rightarrow \tau^+ \tau^- \) decay is not possible in this model. Therefore, observation of \( Z' \rightarrow \tau^+ \tau^- \) decay would rule out our model. Since the \( U(1)_X \) charges for different families of fermions are not diagonal, the gauge interactions of the new gauge boson with the charged fermions are given as :
\begin{equation}\label{flavourgaugeinteraction}
    \mathcal{L}_{int} = g' Z'\left[ 3 \bar{u}_R^{'} \gamma^{\mu} u^{'}_{R} +  \bar{c} ^{'}_R \gamma^{\mu} c^{'}_{R} - 3 \bar{d} ^{'}_R \gamma^{\mu} d^{'}_{R} -
    \bar{s}^{'}_R \gamma^{\mu} s^{'}_{R} - 3 \bar{e}^{'}_R  \gamma^{\mu} e^{'}_{R} 
   - \bar{\mu} ^{'}_R \gamma^{\mu} \mu^{'}_{R} \right] + \textrm{h.c} ~.
\end{equation}
Here, \( u_R', d_R', c_R', s_R', e_R', \mu_R' \) are interaction eigenstates. To obtain their mass eigenstates, we need to rotate them using \( R^u \) and \( R^d \) given in Eq.~\eqref{rotationquarks}. The Flavour changing interaction at leading order is given by :
\begin{multline}\label{flavourmassinteraction}
\mathcal{L}_{int}^{'} = g' Z'\Bigl[\left(s^u_{23} \bar{c}_R \gamma_{\mu} t_R - s^d_{23} \bar{s}\gamma_{\mu} b_R - s^l_{23} \bar{\mu}_R \gamma_{\mu} \tau_R \right) - 4\left(s^d_{12} \bar{d}_R \gamma_{\mu} s_R + s^l_{12}\bar{e}_R \gamma_{\mu} \mu_R \right) \\
+ 4 s^u_{12} \epsilon^2 \bar{c}_R \gamma_{\mu} u_R 
+ \left((3 s^d_{13} - 4 s^d_{12} s^d_{23})\bar{d}_R \gamma_{\mu} b_R + (3 s^l_{13} - 4 s^l_{12} s^l_{23}) \bar{e}_R \gamma_{\mu} \tau_R ) \right)
\Bigr] + \textrm{h.c} ~.
\end{multline}
As a model prediction, we calculate the branching fractions of lepton flavour-violating rare decays \( \mu^- \rightarrow e^- e^- e^+ \) and \( \mu^- \rightarrow e^- \gamma \). In our model, the \( \mu \rightarrow 3e \) decay can occur at tree level through the new gauge bosons, as shown in Fig.~\ref{fig:muondecay}, while the \( \mu \rightarrow e \gamma \) decay is possible at the one-loop level, as shown in Fig.~\ref{fig:mutoeggma}.

The branching ratio of the $\mu \rightarrow 3 e $ decay is obtained as :
\begin{equation}
   Br\left(\mu \rightarrow 3 e \right) = 16 \left(s_{12}^l\right)^2\left(\frac{g' M_W}{g_w M_{Z'}}\right)^4 ~,
\end{equation}
where \( s_{12}^l \) is the mixing angle of the first and second generations of charged leptons, given in Eq.~\eqref{leptonangle}, \( M_W \) is the W-boson mass \( g_w \) is the weak coupling constant, \( M_{Z'} \) is the new gauge boson mass and \( g' \) is the new \( U(1)_X \) gauge coupling constant. Following \cite{Lavoura:2003xp, Cruz-Albaro:2019hrk}, we obtain the branching ratio for $\mu \rightarrow e \gamma $ as :
 \begin{equation}\label{mutoegammabr}
     Br(\mu \to e \gamma) = \frac{\alpha}{2}\left(1-\left(\frac{m_e}{m_{\mu}}\right)^2\right)^3 \Big[\left|\Omega_{\mu l} \Omega_{e l}\right|^2 \left| y_1+y_2+y_3+y_4 \right|^2\Big] \frac{m_\mu}{\Gamma_\mu} ~~,
 \end{equation}
where \( \alpha \) represents the fine structure constant and \( \Gamma_{\mu} \) is the total decay width of the muon. The variables \( y_1, y_2, y_3, \) and \( y_4 \) contain the loop contributions which are explicitly provided in the Appendix~\ref{appendix}. The terms \( \Omega_{\mu l} \) and \( \Omega_{e l} \) denote the coupling strengths of the charged leptons \( l \) with the new gauge boson. These couplings are given by \( \Omega_{\mu e} = 4 g' s^l_{12} \), \( \Omega_{\mu \tau} = g' s^l_{23} \) and \( \Omega_{e \tau} = g' (s^l_{13} - 4 s^l_{12} s^l_{23}) \).

\begin{figure}[ht]
    \centering
    \begin{tabular}{c c}
\includegraphics[width=0.5\linewidth]{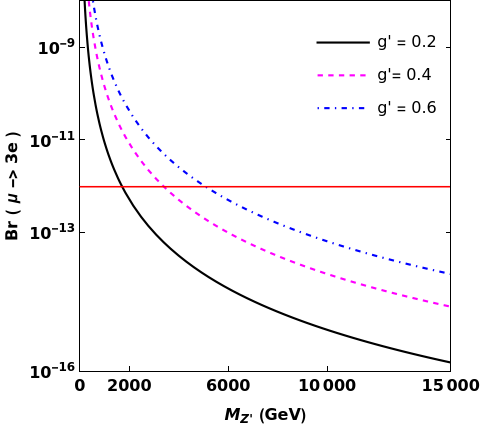} & \includegraphics[width=0.5\linewidth]{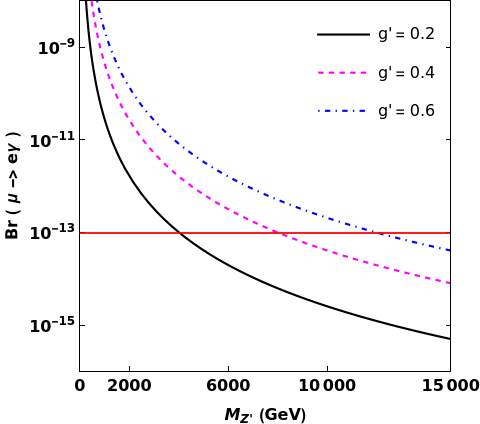} \\
    \end{tabular}
 \caption{Branching ratio of $\mu \to 3 e $ and $\mu \to e \gamma$ for fixed value of gauge coupling constant $g'$. The red line indicate the experimental bound of the branching ratio. The region above the red line are excluded and below the red line are allowed for the new gauge boson mass}
\label{fig:branchingratio}
\end{figure}

The experimental upper bounds for the branching ratios of \( \mu \rightarrow 3 e \) and \( \mu \rightarrow e \gamma \) are \( 1.0 \times 10^{-12} \) and \( 4.2 \times 10^{-13} \), respectively \cite{ParticleDataGroup:2020ssz}. These provide a lower bound on \( M_{Z'} \) for a given value of \( g' \). We have shown the variation of these branching ratios with the new gauge boson mass while keeping the new gauge coupling \( g' \) fixed at \( 0.2 \), \( 0.4 \) and \( 0.6 \) in Fig.~\ref{fig:branchingratio}. Here, we calculate the values of the charged lepton mixing angles by taking the values of the Yukawa couplings and \( \epsilon \)'s as given later in section \ref{benchmark}.

As we can see from the Fig.~\ref{fig:branchingratio}, the \( \mu \to 3 e \) decay provides lower bounds on the \( M_{Z'} \) mass of 1.8, 3.5, and 5.5 TeV, while the \( \mu \to e \gamma \) decay provides lower bounds on the \( M_{Z'} \) mass of 4, 8, and 12 TeV for \( g' = 0.2, 0.4 \) and \( 0.6 \). This means that the \( \mu \to e \gamma \) decay gives a more stringent bound on the \( M_{Z'} \) mass than that obtained from the \( \mu \to 3 e \) decay.

In the quark sector, the FCNC processes are possible due to following interaction terms :
\begin{align}
    (d_R \gamma_{\mu} b_R)^2 + \textrm{h.c} ~, &&(s_R \gamma_{\mu} b_R)^2 + 
    \textrm{h.c} ~, &&
    (d_R \gamma_{\mu} s_R)^2 + \textrm{h.c} 
    ~.
\end{align}
These will contribute to the mixing of \( \bar{B}^0 - B^0 \), \( \bar{B}_s^0 - B_s^0 \), and \( \bar{K}^0 - K^0 \) mixing data. As shown in \cite{Kownacki_2017}, we can find the contribution of these operators to the mass splitting of the various mesons as :
\begin{align}\label{flavouranomalyquark}
     \Delta M_B &= 4.5 \times 10^{-2} (3 s^d_{13}- 4 s_{12}^d s_{23}^d)^2 (g'^2/m_{Z'}^2) ~~\textrm{GeV}^3 ~, \\
     \Delta M_{B_s} &= 6.4 \times 10^{-2}  (s_{23}^d)^2(g'^2/m_{Z'}^2) ~~\textrm{GeV}^3 ~, \\
     \Delta M_K &= 1.9 \times 10^{-3}  (4 s_{12}^d)^2 (g'^2/m_{Z'}^2) ~~\textrm{GeV}^3 ~.
 \end{align}
If we take \( M_{Z'} = 10 \) TeV and \( g' = 0.1 \), then this contribution and the SM contribution can explain the experimental values of these mass splittings \cite{ParticleDataGroup:2020ssz}. For a detailed discussion on FCNCs in family non-universal models, one can see \cite{Chiang:2006we}. 
\section{Renomalizable dimension 4 operator realization}\label{dim4}
We can describe the effective higher-dimension operators of Eq.~\eqref{eftlag} in terms of dimension-four operators by adding some vector-like fermions, as shown in Table~\ref{tab:2}.
\begin{table}[ht]
 \caption {Charges of vector-like fermions}
   \centering
    \begin{tabular}{m{3.5cm} m{2cm} m{2cm} m{2cm} m{2.5cm} m{2.5cm}}
        \hline
        Particles & $SU(3)_C$ & $SU(2)_L$ & $U(1)_Y$ & $U(1)_X$ & $Z_2$ \\
        \hline
        $f_i^u$ &  3 & 1 & 2/3 & (0,1,2) & (+,+,+) \\
        $f_{i}^d$ & 3 & 1 & -1/3 & (0,-1,-2) & (+,+,+)\\
        $f_{i}^l$ & 3 & 1 & -1 & (0, -1,-2) & (+,+,+) \\
       \hline
  \end{tabular}
    \label{tab:2}
\end{table}

\begin{figure}[ht]
    \centering
    \includegraphics[scale=0.80]{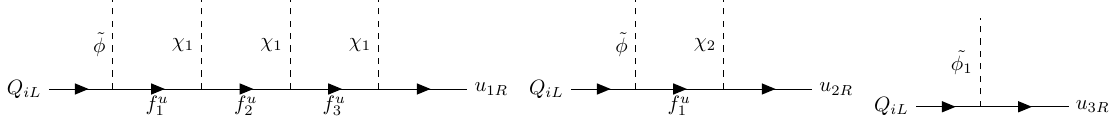}
    \caption{Up type quarks mass generation through dimension four operators}
    \label{fig:dim4up}
\end{figure}

Due to the vector-like nature of the extra fermions, they do not create any anomaly \cite{Peralta:2017qhu,Alves:2023ufm}. With these charge assignments of vector-like fermions, we can write the renormalizable, dimension-4 Yukawa interactions associated with up-type quarks as :
\begin{multline}\label{dim4up}
    \mathcal{L}_{u}^{dim = 4} = \bar{Q}_{iL} \tilde{\phi} f^u_0 + \bar{f}_0 \chi_{1} f^u_1 +\bar{f}^u_1 \chi_{1} f^u_2 + \bar{f}^u_2 \chi_{1} u_{1R} +  \bar{Q}_{iL} \tilde{\phi} f^u_0 + \bar{f}_0 \chi_{1} f^u_1 + \bar{f}^u_1 \chi_1 u_{2R}\\ +\bar{Q}_{iL} \tilde{\phi} ~u_{3R} + \textrm{h.c} 
     ~.
\end{multline}
The Feynman diagram associated with Eq.~\eqref{dim4up} is shown in Fig.~\ref{fig:dim4up}.

Similarly, we can write the renormalizable, dimension-4 Yukawa interactions associated with down-type quarks as :
\begin{multline}\label{dim4down}
    \mathcal{L}_{d}^{dim = 4} = \bar{Q}_{iL} \phi f^d_0+ \bar{f}_0^d \chi^{*}_1 f_1^d +\bar{f}_1^d \chi^*_1 f_2^d + \bar{f}_2^d \chi_2 d_{1R} + \bar{Q}_{iL} \phi f_0^d + \bar{f}_0^d \chi_1^* f_2^d + \bar{f}_2^d \chi_{2} d_{2R} \\
    + \bar{Q}_{iL}^d \phi f_0^d + \bar{f}_0^d \chi_{2} d_{3R} + \textrm{h.c} 
     ~. 
\end{multline}

\begin{figure}[ht]
    \centering
    \includegraphics[scale=0.75]{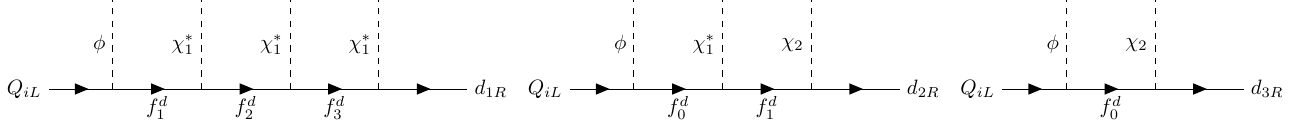}
    \caption{Down type quarks mass generation through dimension four operators}
    \label{fig:dim4down}
\end{figure}

The Feynman diagram associated with Eq.~\eqref{dim4down} is shown in Fig.~\ref{fig:dim4down}.

Integrating out the vector-like fermions in Fig.~\ref{fig:dim4up} and \ref{fig:dim4down} reproduces higher-dimensional operators, of Eq.~\eqref{eftlag}. The renormalizable, dimension-4 Yukawa interactions associated with charged leptons are similar to those of down-type quarks. For this, it is required to replace \( f^d_i \) with \( f^l_i \), \( \bar{Q}_{iL} \) with \( \bar{L}_{iL} \), and \( d_{iR} \) with \( e_{iR} \).
\section{Benchmark value of Yukawa couplings}\label{benchmark}
We perform a precise fit to the observed masses and mixing angles using a cost function defined as :
\begin{equation}
      \chi^2_{O} = \sum_{i} \left( \frac{\langle O_{i} \rangle - \hat{O_i}}{\Delta O_i}\right)^2 ~, 
\end{equation}
where \( O_i \) is the observable with a measured value \( \langle O_i \rangle \pm \Delta O_i \) and a theoretical prediction \( \hat{O_i} \) \cite{Fedele:2020fvh}. For fitting quark masses and mixing, \( i \) runs over six quark masses and three CKM mixing angles, with the cost function denoted by \( \chi^2_{Q} \). For fitting the charged lepton masses, \( i \) runs over the three charged lepton masses, and the cost function is denoted by \( \chi^2_l \). For fitting the neutrino masses and mixing, \( i \) runs over two neutrino mass-squared differences and the PMNS mixing angles, with the cost function denoted by \( \chi^2_{\nu} \). We minimize these cost functions to find the optimal values for the Yukawa couplings that reproduce the observed masses and mixing of the fermions. The masses of the charged fermions and quark mixing angles at \( 1 \, \text{TeV} \) are given in \cite{Xing:2007fb} as :
\begin{align}
      (m_t, m_c, m_u) =& ( 150.7 \pm 3.4,~ 0.532^{+0.074}_{-0.073},~1.10^{+0.43}_{-0.37} \times 10^{-3}) ~~\textrm{GeV} ~, \nonumber \\
      (m_b, m_s, m_d) =& (2.43\pm 0.08,~ 4.7^{+1.4}_{-1.3} \times 10^{-2},~ 2.50^{+1.08}_{-1.03} \times 10^{-3}) ~~\textrm{GeV} ~,
     \nonumber \\
      (m_{\tau}, m_{\mu}, m_e) =& (1.78\pm 0.2,~ 0.105^{+9.4 \times 10^{-9}}_{-9.3 \times 10^{-9}},~ 4.96\pm 0.00000043 \times 10^{-4}) ~~\textrm{GeV} ~,
\end{align}
\begin{align}
     \sin \theta_{12}^q &=0.22650 \pm 0.00048 ~, &
    \sin \theta_{23}^q &=0.04053 \pm 0.00072 ~,
    & \sin \theta_{13}^q &= 0.00361\pm 0.0007 ~.
\end{align}
To calculate the values of the Yukawa couplings, we minimize these cost functions by considering $\epsilon = \epsilon' = 0.0236$ and varying the Yukawa couplings in the range of $0.1$ to $4$. We obtain minimum values of $\chi_Q^2$ and $\chi_l^2$ as 3.29 and 0.11, respectively and the corresponding Yukawa couplings for charged fermions are given by :
\begin{align}
    h^{u} &=
    \begin{pmatrix}
        0.10 & 0.30 & 0.46\\
        3.35 & 0.10 & 0.35 \\
        0.11 & 0.44 & 0.64
    \end{pmatrix} ~,
&
h^{d} &=
    \begin{pmatrix}
        3.99 & 0.16 & 0.19\\
        1.10 & 1.39 & 0.50\\
        4.00 & 0.13 & 0.18
    \end{pmatrix}~,
&
h^{l} &=
    \begin{pmatrix}
        0.10 & 1.06 & 0.10\\
        0.24 & 0.10 & 0.38\\
        0.10 & 0.59 & 0.17
    \end{pmatrix}~~.
\end{align}
For the above values of Yukawa couplings, the masses of the charged fermions and the mixing angles of the quarks are as follows :
\begin{align}
    (m_t, m_c, m_u) &\approx ( 150.57 , 0.516673, 1.10109 \times 10^{-3}) ~~ \text{GeV} ~,\nonumber \\
(m_b, m_s, m_d) &\approx (2.34, 4.52 \times 10^{-2}, 2.50 \times 10^{-3}) ~~ \text{GeV} ~,
\nonumber \\
(m_{\tau}, m_{\mu}, m_e) &\approx (1.78 , 0.105, 4.96 \times 10^{-4}) ~~ \text{GeV} ~,
\end{align}
\begin{align}
     \sin \theta_{12}^q &= 0.226456 ~, &
    \sin \theta_{23}^q &= 0.0405297 ~,&
    \sin \theta_{13}^q &=  0.00353272 ~~.
\end{align}
As we can see, these values for quark masses, mixing, and charged lepton masses provide a good fit within the experimental values of these quantities. Now, the mass-squared differences and mixing angles of neutrinos for the normal hierarchy are given in \cite{ParticleDataGroup:2020ssz} as :
\begin{align}
    \Delta m_{21}^2 &= (7.53 \pm 0.18) \times 10^{-5} ~~ {\text{eV}}^2 ~, &
    \Delta m_{32}^2 &= (2.453 \pm 0.033) \times 10^{-3} ~~ {\text{eV}}^2  ~,
\end{align}
\begin{align}
     \sin^2\theta_{12}^{\nu} &= 0.307 \pm 0.013 ~, &
    \sin^2\theta_{23}^{\nu} &= 0.546 \pm 0.021 ~, &
    \sin^2\theta_{13}^{\nu} &= (2.20 \pm 0.07) \times 10^{-2} ~.
\end{align}
These values are the same for the inverted hierarchy, except for :
\begin{align}
    \Delta m_{32}^2 &= (-2.536 \pm 0.034) \times 10^{-3} ~~ {\text{eV}}^2 ~, &
    \sin^2\theta_{23}^{\nu} &= 0.539 \pm 0.021 ~.
\end{align}
For our numerical analysis, we take \( v_{\eta} = 1 \, \text{keV} \) and minimize \( \chi_{\nu}^2 \) to obtain the neutrino Yukawa couplings for both normal and inverted hierarchy cases. We find the minimum values of \( \chi^2_{\nu} \) for the normal hierarchy and inverted hierarchy cases to be 0.49 and 2.13, respectively, with the Yukawa couplings for the normal and inverted hierarchies given by :
\begin{equation}
(h^{\nu}_{11}, h^{\nu}_{22}, h^{\nu}_{33}, h^{m}_{11})_{\text{Normal}} = (0.52, 1.27, 2.85, 0.92) ~,
\end{equation}
\begin{equation}
(h^{\nu}_{11}, h^{\nu}_{22}, h^{\nu}_{33}, h^{m}_{11})_{\text{Inverted}} = (0.47, 0.20, 2.55, 0.59) ~.
\end{equation}
All other neutrino Yukawa couplings are equal to 1, as mentioned in Section \ref{neutrino}.
The neutrino mass squared differences and mixing angles for the values of the Yukawa couplings obtained for fitting the normal hierarchy are :
\begin{align}
    \Delta m_{21}^2 &= 7.5293 \times 10^{-5} ~~ {\text{eV}}^2 ~, &
    \Delta m_{32}^2 &= 2.45244 \times 10^{-3} ~~ {\text{eV}}^2 ~,
\end{align}
\begin{align}
     \sin^2\theta_{12}^{\nu} &= 0.313472 ~, &
    \sin^2\theta_{23}^{\nu} &= 0.535814 ~, &
    \sin^2\theta_{13}^{\nu} &= 2.19401 \times 10^{-2} ~.
\end{align}
Neutrino mass squared differences and mixing angles for the values of the Yukawa couplings obtained for fitting the inverted hierarchy are :
\begin{align}
    \Delta m_{21}^2 &= 7.53002 \times 10^{-5} ~~ {\text{eV}}^2 ~,&
    \Delta m_{32}^2 &= -2.53577 \times 10^{-3} ~~ {\text{eV}}^2 ~.
\end{align}
\begin{align}
    \sin^2\theta_{12}^{\nu} &= 0.320691~, &
    \sin^2\theta_{23}^{\nu} &= 0.560994~, &
    \sin^2\theta_{13}^{\nu} &= 2.21119 \times 10^{-2} ~.
\end{align}
These values of the neutrino mass-squared differences and mixing angles provide a good fit within the experimental values of these quantities in both cases of normal and inverted hierarchy.
\section{Conclusion}\label{conclusion}
By extending the SM gauge symmetry with additional \( U(1)_X \) and \( Z_2 \) symmetries, we have successfully explained the fermion masses and mixing hierarchies, including neutrino masses and mixings, by introducing three right-handed neutrinos, two scalar singlets, and a new scalar doublet, with \( O(1) \) Yukawa coupling values. Both the normal and inverted hierarchy scenarios can be explained with \( O(1) \) Yukawa couplings in this model. Since the heavy right-handed neutrinos and the new scalars are at the TeV scale, signatures of these particles may be observable in near-future experiments. 

As a prediction of our model, we have calculated the branching ratios for two lepton flavour-violating processes, \( \mu \rightarrow 3e \) and \( \mu \rightarrow e \gamma \), finding that the \( \mu \rightarrow e \gamma \) process provides a more stringent lower bound on the new gauge boson mass for given values of the gauge coupling constant. 

We also discussed a possible UV completion of the theory by introducing extra vector-like fermions, where all interactions are expressed by renormalizable dimension-4 operators.
\section*{Acknowledgement}
ARS thanks the Ministry of Minority Affairs, Government of India, for financial support through Maulana Azad National Fellowship (No. F. 82-27/2019 (SA-III)). 
\appendix 
\section{Expressions of Loop Coefficients} \label{appendix}
Different $y_i$'s in Eq.~\eqref{mutoegammabr} are given below :
\begin{eqnarray*}
     y_1 &=& m_{\mu}\left[2 a + 6 b + 3 c+ \left(\frac{m_l}{m_Z'}\right)^2 \left(\frac{3 c}{2} - b \right) + \left(\frac{m_e}{m_{Z'}}\right)^2 \left(\frac{3 c}{2} + b \right)\right] ~,\\
     y_2 &=& m_{e}\left[2 a + 6 b + 3 c+ \left(\frac{m_l}{m_Z'}\right)^2 \left(\frac{3 c}{2} - b \right) + \left(\frac{m_{\mu}}{m_{Z'}}\right)^2 \left(\frac{3 c}{2} + b \right)\right] ~,\\
      y_3 &=& m_{l}\left[-4 a - 8 b + 2 c \left(\frac{m_l}{m_Z'}\right)^2 - \left(\frac{m_{\mu}}{m_{Z'}}\right)^2 \left(\frac{3 c}{2} + b \right) - \left(\frac{m_{e}}{m_Z'}\right)^2 \left(\frac{3 c}{2} + b \right)\right] ~,
\end{eqnarray*}
\begin{equation}
    y_4 = - \frac{ 3  \, c \, m_e m_{\mu} m_{l}}{m_{Z'}^2} ~,
\end{equation}
where
\begin{eqnarray}
   a &=& \frac{\textrm{i}}{16 \pi^2 m_{Z'}^2}\left[-\frac{1}{t-1}+\frac{\ln t}{(t-1)^2}\right] ~, \nonumber \\  
   b &=& \frac{\textrm{i}}{16 \pi^2 m_{Z'}^2}\left[ \frac{t-3}{4(t-1)^2}+\frac{\ln t}{2(t-1)^3}\right] ~,  \nonumber \\ 
    c &=& \frac{\textrm{i}}{16 \pi^2 m_{Z'}^2}\left[ \frac{-2 t^2+7 t -11}{1`8(t-1)^3}+\frac{\ln t}{3(t-1)^4}\right]~,
\end{eqnarray}
with 
\begin{equation}
    t = \left(\frac{m_l}{m_{Z'}}\right)^2 ~.
\end{equation}
\bibliographystyle{JHEP}
\bibliography{flavour.bib}
\end{document}